\documentclass[11pt,twoside]{article}
\usepackage{./asp2014}

\aspSuppressVolSlug
\resetcounters

\begin{document}

\title{Multi-Object Spectroscopy with MUSE}
\author{Andreas Kelz$^1$, Sebastian Kamann$^2$, Tanya Urrutia$^1$, Peter Weilbacher$^1$, Lutz Wisotzki$^1$ and Roland Bacon$^3$ 
\affil{$^1$Leibniz-Institut f\"ur Astrophysik Potsdam (AIP), An der Sternwarte 16, D-14482 Potsdam, Germany; \email{akelz@aip.de}}
\affil{$^2$Institut f\"ur Astrophysik, Universit\"at G\"ottingen (IAG), Friedrich-Hund-Platz 1, D-37077 G\"ottingen, Germany; \email{sebastian.kamann@phys.uni-goettingen.de}}
\affil{$^3$CRAL, Observatoire de Lyon, CNRS, Universit\'{e} Lyon 1, 9 Avenue Ch. Andr\'{e} F-69561 Saint Genis Laval Cedex, France; \email{roland.bacon@univ-lyon1.fr}}
}

% This section is for ADS Processing.  There must be one line per author.
\paperauthor{Andreas Kelz}{akelz@aip.de}{ }{Leibniz-Institut f\"ur Astrophysik Potsdam (AIP)}{3D \& Multi-Object Spectroscopy}{Potsdam}{Brandenburg}{14482}{Germany} 
%\paperauthor{Sample~Author1}{Author1Email@email.edu}{ORCID_Or_Blank}{Author1 Institution}{Author1 Department}{City}{State/Province}{Postal Code}{Country}
%\paperauthor{Sample~Author2}{Author2Email@email.edu}{ORCID_Or_Blank}{Author2 Institution}{Author2 Department}{City}{State/Province}{Postal Code}{Country}

\begin{abstract}
Since 2014, MUSE, the Multi-Unit Spectroscopic Explorer, is in operation at the ESO-VLT. 
It combines a superb spatial sampling with a large wavelength coverage.  
By design, MUSE is an integral-field instrument, but its field-of-view and large multiplex 
make it a powerful tool for multi-object spectroscopy too. 
Every data-cube consists of 90,000 image-sliced spectra and 3700 monochromatic images. 
In autumn 2014, the observing programs with MUSE have commenced, with targets ranging from distant galaxies to local stellar populations, star formation regions and globular clusters. 

This paper provides a brief summary of the key features of the MUSE instrument 
and its complex data reduction software. 
Some selected examples are given, how multi-object spectroscopy for hundreds of continuum and emission-line objects can be obtained in wide, deep and crowded fields with MUSE, without the classical need for any target pre-selection.   
\end{abstract}

\section{The MUSE instrument in brief}
The Multi Unit Spectroscopic Explorer (MUSE) \citep{bacon_2014} is a 3D-Spectrograph
at the Very Large Telescope (VLT) of the European Southern Observatory (ESO). 
It combines a 1 square arcminute field-of-view with a seeing-limited spatial sampling, resulting in 90,000 spectra per exposure. After 10 years of development, assembly, and testing, MUSE was commissioned at the Paranal Observatory in 2014. Meanwhile the science verification and the initial observing runs have yielded excellent results, demonstrating the power of this new facility for the European astronomical community. 

Currently, MUSE is in operation in the seeing-limited, wide-field-mode (WFM). In the future, ESO will upgrade the UT4 with the VLT Adaptive Optics Facility (AOF). Together with the GALACSI-AO system, this will yield an improved spatial resolution, even higher sensitivity and enable the AO-assisted narrow-field-mode (NFM) for MUSE. Table 1 summarizes the top instrumental parameters and observing modes.  

\begin{table}[!ht]
\caption{MUSE instrumental parameters in a nutshell}
\smallskip
\begin{center}
{\small
\begin{tabular}{ll} 
\tableline
\noalign{\smallskip}
Parameter & Value \\
\noalign{\smallskip}
\tableline
\noalign{\smallskip}
Number of IFU modules: 	& 	24 (image slicer + spectrograph + CCD) \\ 
Wavelength coverage: 	& 	480-930 nm (nominal range) \\ 
~ 			& 	465-930 nm (extended range) \\ 
Throughput WFM: 	&	14\% (at 480 nm), 35\% (at 750 nm), 14\% (at 930 nm) \\ 
Field of View: 		&	59" $\times$ 60" (in WFM) and 7".5 $\times$ 7".5 (in NFM) \\ 
Spatial sampling: 	&	0".2 (in WFM) and 0".025 (in NFM) \\  
Multiplex factors: 	& 	1152 slices, 90,000 spaxels, 3700 wavelength bins \\ 
\noalign{\smallskip}
\tableline 
\end{tabular}
}
\end{center}
\end{table}

The MUSE data reduction software \citep{weilbacher_2014} converts and calibrates the raw data from the 24 CCDs into a combined data-cube (with two spatial and one wavelength axis) and corrects for instrumental and atmospheric effects. Apart from standard procedures such as bias subtraction, cosmics removal, spectra extraction, etc, it reconstructs the image at the focal plane from the 1152 slices. A special geometrical calibration was performed to measure the edges of each projected image slicer in the field. The software also features methods for sky subtraction and flux calibration of the data. In addition to each data cube, a variance cube is being generated. The pipeline was operational at the time of MUSE first light and was a crucial contribution for the successful commissioning. The latest release of the pipeline to be used with the \mbox{EsoRex} program can be downloaded from the ESO site ({\url{www.eso.org/sci/software/pipelines/muse/muse-pipe-recipes.html}}).

\section{The multiplex advantage} 

\vspace{-5mm}
\articlefigure[width=0.9\textwidth]{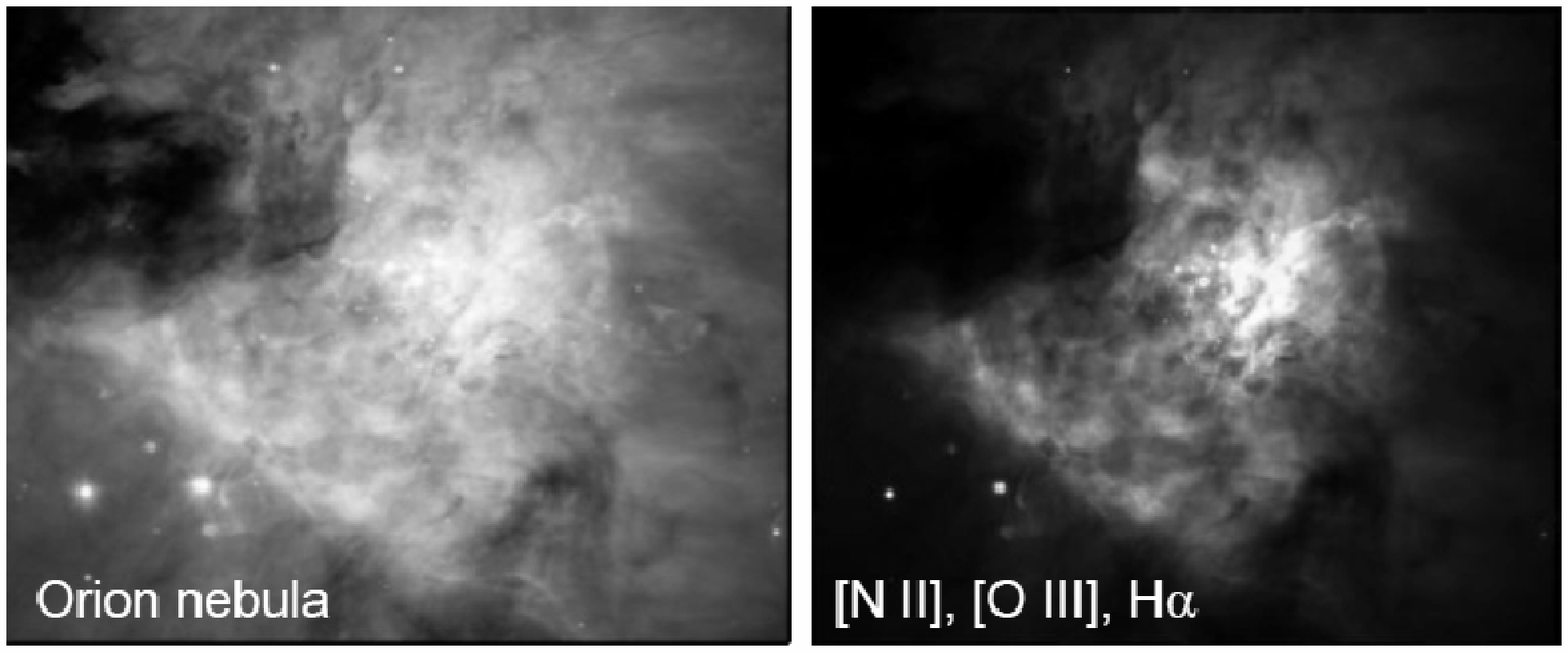}{fig1}{This panoramic view of the Orion nebula (M~42) was taken with MUSE as a mosaic of 5$\times$6=30 individual pointings. The entire data-cube (size: 110 Gb) contains 2.7 million spectra ranging from 460 to 930~nm. \emph{Left:} A broad-band composite, extracted from spectral regions. \emph{Right:} A different view, derived from the very same data-cube, but using the fluxes from three emission lines instead.}  

Contrary to multi-slit or multi-fiber spectrographs, MUSE uses the entire FoV to obtain spectra for each spaxel. To demonstrate the multiplex power of MUSE, to verify its performance, and to stress-test the data reduction pipeline, a mosaic of 30 pointings was obtained in the center of the Orion nebula (see Fig. \ref{fig1}). The resulting data-cube allowed the detection and spectral analysis of point-like and extended sources, as well as the creation of flux, line ratio and velocity maps of the entire region \citep{weilbacher_2015}.

\subsection{MUSE in deep fields} 

A core science case in the development of MUSE has been the observation of high red-shifted galaxies in the early universe to study galaxy formation. During the final commissioning run, the consortium observed the Hubble Deep Field South (HDFS) for 27 hours integration time \citep{bacon_2015} resulting in an unprecedented depth. 
The redshifts of 189 sources were measured, increasing by more than an order of magnitude
the number of known spectroscopic redshifts in this field. In addition, 26 Ly-$\alpha$ emitting galaxies were discovered, which are not detected in the HST WFPC2 broad band images (see Fig. \ref{fig23} left). 

The entire data set, including the data cube, associated variance cube, and the source catalog with redshifts, spectra and emission line fluxes, was released to the community to allow for follow-up studies ({\url{http://muse-vlt.eu/science/}}). 

\articlefiguretwo{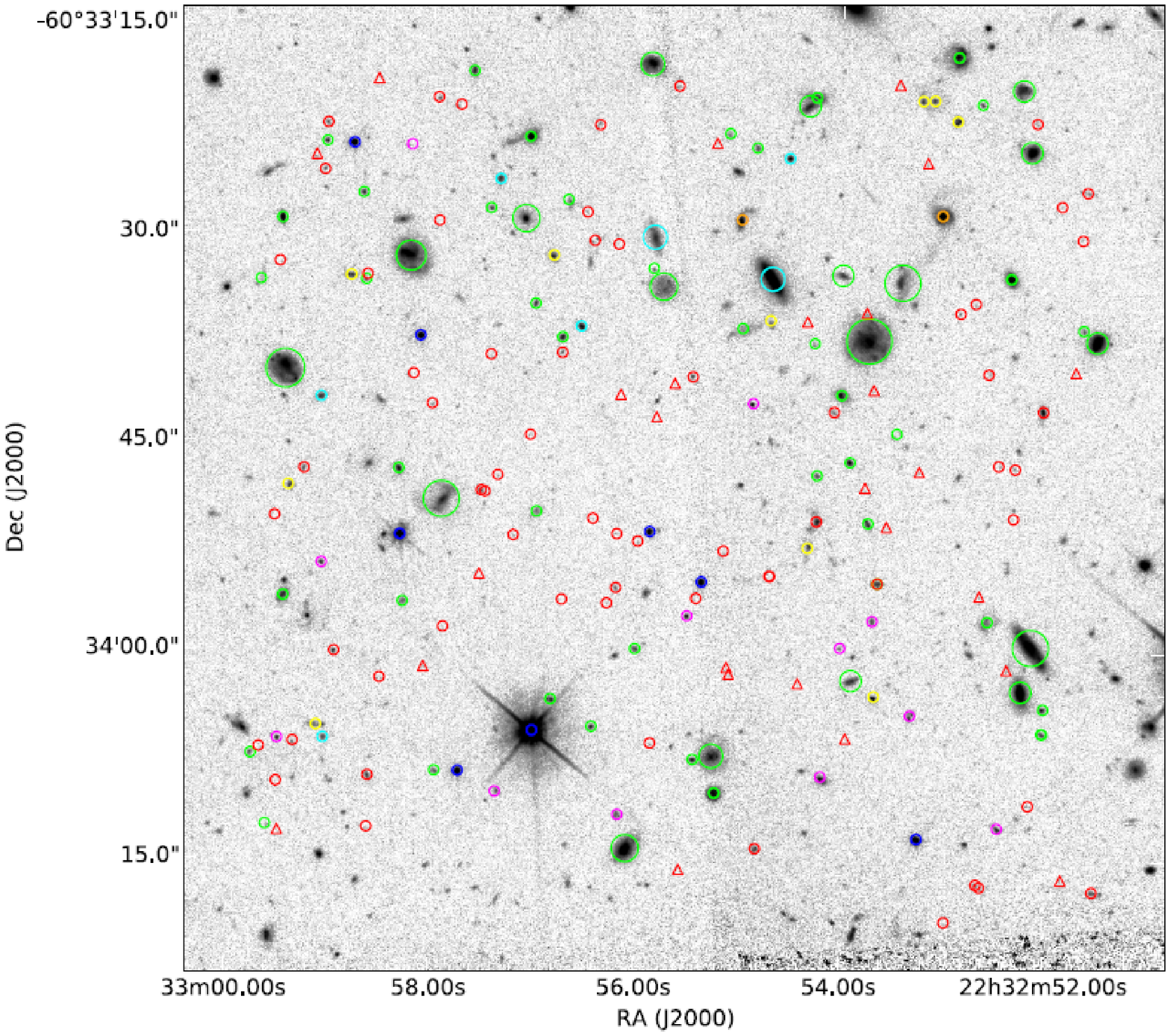}{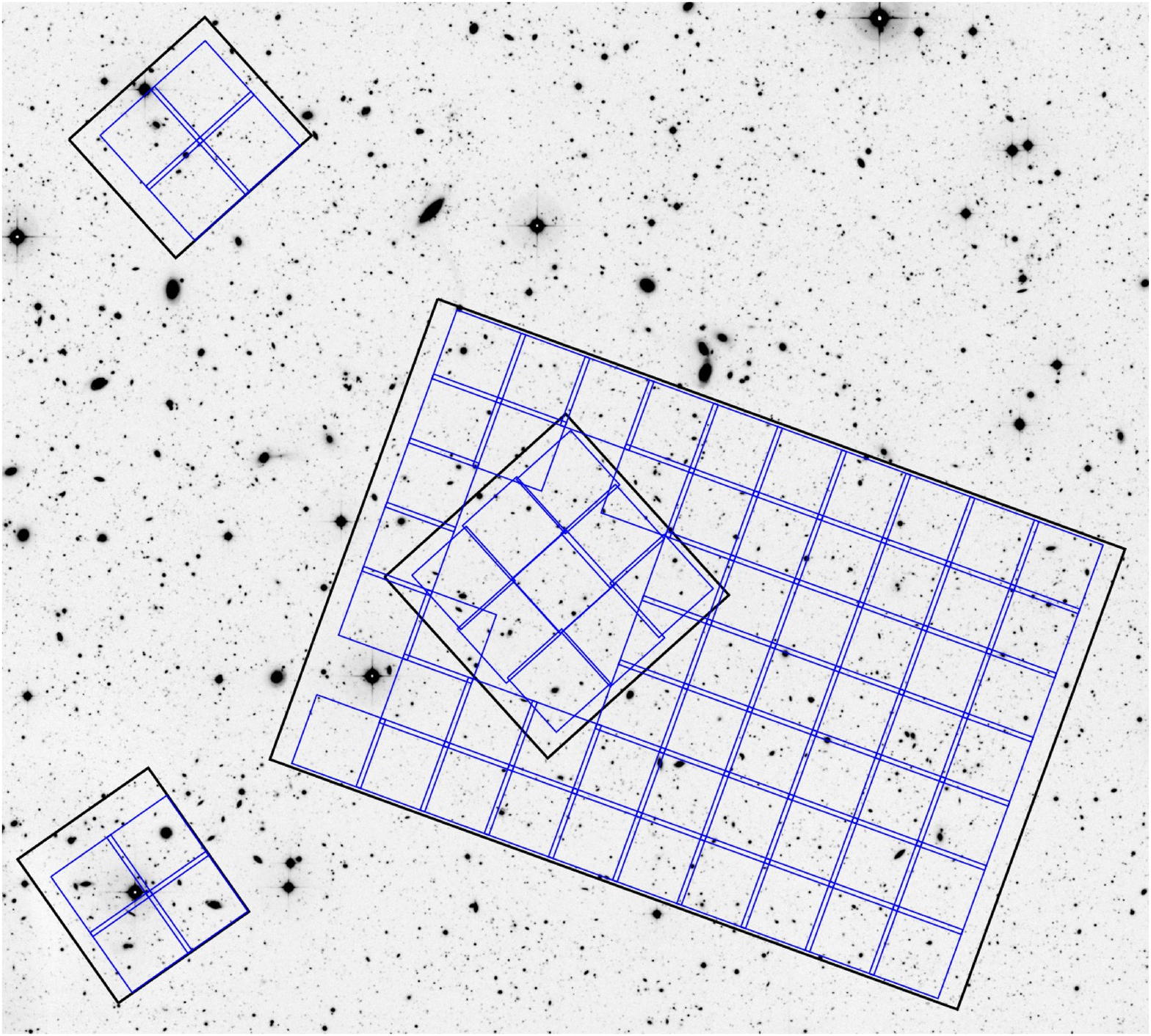}{fig23}
{\emph{Left:} Marked with symbols on the HDFS are the 181 galaxies with secure redshift, resulting from 27 hours of MUSE integration. The triangles indicate the 26 faint (R>29.5) Ly-$\alpha$ emitter galaxies detected by MUSE, but not present in the HST deep WFPC2 images (see \citet{bacon_2015} and eso 1507 Science Release). 
\emph{Right:} Field selections for the MUSE wide and deep surveys in the UDF/GOODS-S region. With a mosaic of overlapping fields (small squares), the MUSE-wide pointings cover the entire CANDLES-DEEP field (outer rectangle). The Hubble Ultra Deep Field (rotated square) is targeted by the MUSE medium-deep survey.}

\subsection{MUSE in wide fields} 

In a complementary program to the deep-fields, MUSE is used for a wider, but shallower survey over $\sim$100 arcmin$^2$ to perform a census of bright Ly$\alpha$ emitters (Wisotzki et al. in prep.). This MUSE-Wide Survey currently focuses on the CANDELS-DEEP field in the CDFS area, surrounding the Hubble Ultra Deep Field (see Fig. \ref{fig23} right). 
This comprises the most studied area in the sky with excellent multi-wavelength coverage, 
such as deep HST optical (GOODS-South), Chandra, Spitzer, Herschel and ALMA observations. 
While other MOS surveys, such as VANDELS and the VVDS-Deep and deep 3D-HST grism data, have
focused on identifying the galaxy population, MUSE's capabilities lie in the
serendipitous detection of emission line sources, whose continuum is too 
faint to be targeted for follow-up with classical MOS surveys. 
The MUSE-wide survey observes these fields with 1 hour exposure time each, yielding objects down to 24th magnitude in H.

\subsection{MUSE in crowded fields} 

\articlefigure[width=0.75\textwidth]{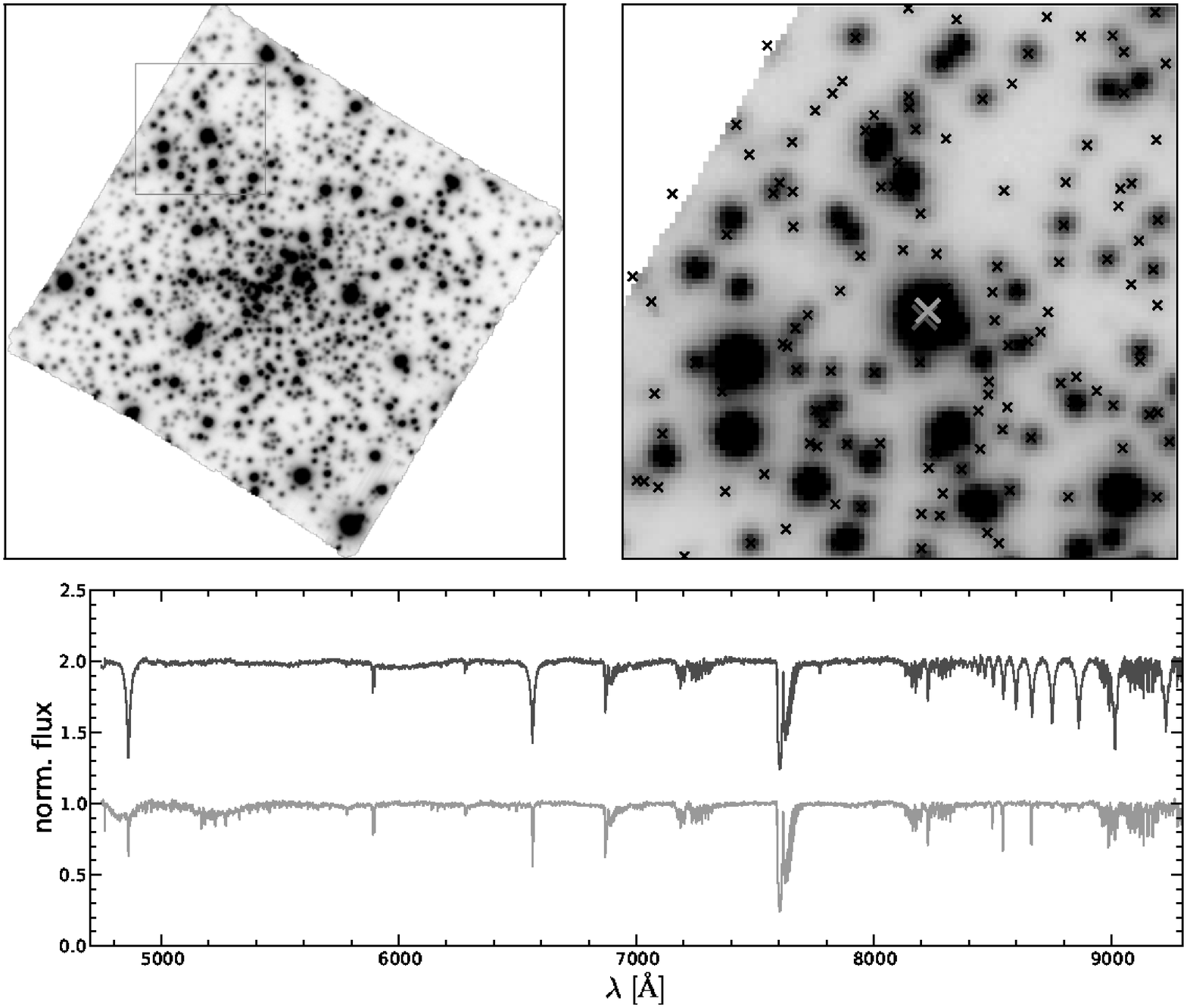}{fig4}{\emph{Top:} Reconstructed MUSE image of the globular cluster NGC~6397 with identified sources (crosses). \emph{Bottom:} By applying the novel technique of crowded field 3D-spectroscopy \citep{kamann_2014}, deblended spectra of up to 5000 stars from a single data-cube were extracted, even of overlapping sources.}

In heavily crowded fields, slit- or fiber-based spectroscopy suffers from the blending of sources and the difficulty to fully record the 2-dimensional area around an object and therefore to disentangle the overlapping light. Combining 3D-spectroscopy with high-resolution imaging data, a technique was developed \citep{kamann_2013} to optimize the deblending and extraction of spectra in very crowded
fields by applying point spread function fitting techniques to data-cubes. 
A MUSE observing program targets globular clusters (Fig. \ref{fig4}) to obtain the largest spectroscopic samples in these objects to date \citep{husser_2015}. This data enables a range of applications, ranging from stellar parameter studies to the search for intermediate-black holes in the cluster centers.

%\clearpage 
\acknowledgements MUSE is built by a European-wide consortium of six institutes and ESO, led by CRAL. AIP and IAG gratefully acknowledge support by the BMBF-Verbundforschung (grant no. 05A14BAC and 05A14MGA).

% For non-BibTex:

\end{document}